\documentclass[journal]{IEEEtran}
\usepackage[noend]{algpseudocode}
\usepackage{cite}
\usepackage{amsmath,amssymb,amsfonts}
\usepackage{graphicx}
\usepackage{textcomp}
\usepackage{subfigure}
\usepackage{multirow}
\usepackage{color}
\usepackage{diagbox}

\newtheorem{theorem}{Theorem}{}
\newtheorem{lemma}{Lemma}{}
\newtheorem{assumption}{Assumption}
{}
{}
{}
{}
{}
{}
{}

\begin{document}
\title{A Novel Dynamic Event-triggered Mechanism for Dynamic Average Consensus}
\author{Tao Xu, Zhisheng Duan, Guanghui Wen, and~Zhiyong Sun
\thanks{This work is supported by the National Natural Science Foundation of China under grants T2121002 and 62173006, and the China Postdoctoral Science Foundation under grants 2022TQ0029 and 2022M720435. (Corresponding author: Zhisheng Duan.)}
\thanks{T. Xu is with the School of Mechatronical Engineering,
Beijing Institute of Technology, Beijing, China, and also with the Yangtze Delta Region Academy
of Beijing Institute of Technology, Jiaxing, Zhejiang, China (e-mail: xutao0429@pku.edu.cn).}
\thanks{Z. Duan is with the Department of Mechanics and Engineering Science, Peking University, Beijing, China (e-mail: duanzs@pku.edu.cn).}
\thanks{G. Wen is with the Department of Systems Science, Southeast University, Nanjing, China (e-mail: wenguanghui@gmail.com).}
\thanks{Z. Sun is with the Department of Electrical Engineering, Eindhoven University of Technology, Eindhoven, the Netherlands (e-mail: sun.zhiyong.cn@gmail.com).}
}
\maketitle
\begin{abstract}
  This paper studies a challenging issue introduced in a recent survey, namely designing a distributed event-based scheme to solve the dynamic average consensus (DAC) problem.
First, a robust adaptive distributed event-based DAC algorithm is designed without imposing specific initialization criteria to perform estimation task under intermittent communication.
Second, a novel adaptive distributed dynamic event-triggered mechanism is proposed to determine the triggering time when neighboring agents broadcast information to each other. Compared to the existing event-triggered mechanisms, the novelty of the proposed dynamic event-triggered mechanism lies in that it guarantees the existence of a
positive and uniform minimum inter-event interval without sacrificing any accuracy of the estimation, which is much more practical than only ensuring the exclusion of the Zeno behavior or the boundedness of the estimation error.
Third, a composite adaptive law is developed to update the adaptive gain employed in the distributed event-based DAC algorithm and dynamic event-triggered mechanism. Using the composite adaptive update law, the distributed event-based solution proposed in our work is implemented without requiring any global information.
Finally, numerical simulations are provided to
illustrate the effectiveness of the theoretical results.
\end{abstract}
\section{Introduction}\label{sec1}
The dynamic average consensus (DAC) problem was first formulated and studied in \cite{Spanos2005ifacdynamic}, where each agent without a particular model is assigned a time-varying reference signal and expected to estimate the average of all reference signals.
The DAC problem has gained enormous popularity, due to its numerous applications, such as distributed formation control \cite{Fax2004tacinformation}, distributed state estimation \cite{Olfati2005cdcdistributed}, distributed convex optimization \cite{Damiano2016tacnewton}, distributed resource allocation \cite{Cherukuri2016autoinitialization}, and so forth \cite{Kia2019csmtutorial}.
Many efficient distributed DAC algorithms have been proposed; see \cite{Freeman2006cdcstability, Chenfei2012tacdistributed, Shahram2012autodynamic, Kia2015ijrncdynamic, Chenfei2015tacdistributed, Moradian2017accdynamic, Zhao2017autodistributed} to cite a few.

The distributed DAC algorithms designed in \cite{Freeman2006cdcstability, Chenfei2012tacdistributed, Shahram2012autodynamic, Kia2015ijrncdynamic, Chenfei2015tacdistributed, Moradian2017accdynamic, Zhao2017autodistributed} lack robustness to initialization, since their implementation is dependent on some specific initial conditions.
For example,
the sum of the initial values of the internal states of all agents in these algorithms
are required to be zero.
This condition is violated
when a group of agents is divided into several subgroups due to communication link break. In this case, re-initialization of the algorithm for each subgroup is required
after every network disruption. However, with a specific initialization, one would need to detect the communication failure globally to perform re-initialization, which is impractical in real applications.
To eliminate special initialization requirement, robust distributed DAC algorithms were designed in \cite{George2017accrobust, Xukedong2021tcsiirobust}, where some global information about the communication topology is needed to choose a constant gain.
To reduce dependence on global information, adaptive gains
were introduced in \cite{George2019tacrobust}, resulting in robust adaptive distributed DAC algorithms.

Another issue with the distributed DAC algorithms proposed in \cite{Freeman2006cdcstability, Chenfei2012tacdistributed, Shahram2012autodynamic, Kia2015ijrncdynamic, Chenfei2015tacdistributed, Moradian2017accdynamic, Zhao2017autodistributed} is that continuous-time communication among agents is required. Specifically, the information of each agent is always transmitted to neighbors without a break via wireless communication network, and then calculated in real time through data processing units. This is impractical since the channel bandwidth of wireless communication network and the computing resource of data processing units
are limited \cite{Nowzari2019autoevent}.
An effective approach to reduce the requirement of continuous communication is to take advantage of an event-triggered mechanism, which enables agent-to-agent information sharing to only occur when a specific event is triggered. To date, event-triggered mechanism has been widely utilized in some coordination control scenarios \cite{Cheng2019tacfully, Li2020autoconsensus, Qian2020cyberdistributed}.
Designing distributed event-based solutions to the DAC problem has been regarded as an interesting yet difficult problem in the survey \cite{Nowzari2019autoevent}.
Some efforts have been made to solve this problem; see \cite{Kia2014cdcdynamic, Kia2015autodistributed, George2018cdcdistributed, Yu2021itnecdistributed, Xing2020tcnsrobust}, where the estimation error is zero in \cite{George2018cdcdistributed, Yu2021itnecdistributed} and bounded in \cite{Kia2014cdcdynamic, Kia2015autodistributed, Xing2020tcnsrobust}.

Despite the progress achieved in the references above, a couple of challenging issues still exist and deserve further exploration.
The distributed event-based DAC algorithms developed in \cite{Kia2014cdcdynamic, Kia2015autodistributed} depend on specific initial conditions.
Moreover, some global information, e.g., the minimum nonzero eigenvalue of the Laplacian matrix associated with the global communication topology or the total number of all agents, is required in \cite{Yu2021itnecdistributed, George2018cdcdistributed} to design DAC algorithms or event-triggered mechanisms.
Although a robust adaptive event-based distributed DAC algorithm has been proposed in \cite{George2018cdcdistributed},
the designed adaptive gain has no tendency to decrease, since only nonnegative direct adaptive component is employed to design the standard adaptive update law.
Fast adaptation achieved by the nonnegative standard adaptive update law often accelerates the convergence at the expense of generating high-frequency oscillations of system response \cite{Yucelen2013taclow} and high amplitude of system input \cite{Mei2016tacdistributed}.
To
avoid these implementation issues,
a common method is to make use of the $\sigma$-modification technique \cite{Cheng2019tacfully, Li2020autoconsensus, Qian2020cyberdistributed}, i.e., introducing an additional damping term into the standard adaptive update law. However,
directly following the design in \cite{Cheng2019tacfully, Li2020autoconsensus, Qian2020cyberdistributed}
cannot guarantee a zero estimation error in the DAC problem. Therefore,
under the premise of guaranteeing a zero-converging estimation error,
the design of a robust distributed solution with adaptive gain and
proper adaptive update law is worthy of further investigation.

The triggering functions proposed in \cite{George2018cdcdistributed, Yu2021itnecdistributed} are guaranteed to generate
 a finite number of events within a finite time, namely excluding the Zeno behavior. Nevertheless,
the minimum inter-event interval of each agent tends to zero with time.
It should be pointed out that with faster and faster samplings, information updates and exchanges cannot be implemented on a digital platform \cite{Heemels2012cdcan}.
Therefore, the existence of a positive lower bound of the inter-event intervals is more desirable in practice.
Although a positive minimum inter-event interval is achieved in \cite{Kia2014cdcdynamic, Kia2015autodistributed, Xing2020tcnsrobust} by introducing a positive constant term in the triggering function, the estimation error is only guaranteed to be bounded rather than converging to zero.
To the best of our knowledge, how to design an event-triggered mechanism that guarantees a positive and uniform minimum inter-event interval without sacrificing DAC convergence is still open.

This paper presents for the first time a robust adaptive distributed
event-based solution to deal with the DAC problem with the challenging issues listed above. Compared with the most relevant results \cite{George2018cdcdistributed, Yu2021itnecdistributed, Kia2014cdcdynamic, Kia2015autodistributed, Xing2020tcnsrobust}, the main contributions of our work are summarized in the following:
\begin{enumerate}
  \item Compared with \cite{Kia2014cdcdynamic, Kia2015autodistributed}, the distributed event-based DAC algorithm of our work is robust to initialization, and hence
is applicable to the case of communication link failure.
Different from \cite{Yu2021itnecdistributed, George2018cdcdistributed}, the distributed
event-based solution of our work is performed without involving any global information.
Unlike the standard adaptive update law that only involves a direct adaptive component as in \cite{George2018cdcdistributed},
a $\sigma$-modification component is introduced in our work. In addition to a damping term, a dynamically
updated compensation term is supplemented in the $\sigma$-modification component, which ensures the achievement of zero estimation error.
  \item Differing from the triggering function with a constant term as in \cite{Kia2014cdcdynamic, Kia2015autodistributed, Xing2020tcnsrobust}, which results in a bounded estimation error, the dynamic triggering function proposed in our work guarantees a zero estimation error.
Compared with the triggering function with an exponential decay term or with a dynamically updated internal variable as in \cite{George2018cdcdistributed, Yu2021itnecdistributed}, which only ensures a positive lower bound of the inter-event intervals within a finite time,
the dynamic triggering function proposed in our work guarantees
a positive and uniform lower bound of the inter-event intervals for all time.
\end{enumerate}
\section{Preliminaries}\label{sec2}
\subsection{Notation}\label{sec2.1}
Let $\mathbb{R}$, $\mathbb{R}^{n}$  and $\mathbb{R}^{n\times m}$ denote the sets of real numbers, $n\times1$ real vectors, and $n\times m$ real matrices, respectively.
For a scalar function $f: [0, \infty)\rightarrow\mathbb{R}$, $f(t^{+})$ is denoted as the right-hand limit of $f$ at $t$.
The symbol $A^{\dagger}$ represents the Moore-Penrose inverse of a real matrix $A$.
The notation $\|\cdot\|$ represents the Euclidean vector norm for a vector and the induced matrix norm for a matrix.
The symbol $\mathbf{1}_{n}$ denotes a $n$-dimensional vector whose components are all ones.
For a vector function $g: [0, \infty)\rightarrow\mathbb{R}^{n}$, we say that $g\in\mathrm{L}_{p}, 1\leq p<\infty$, if $(\int_0^\infty \|g(t)\|^{p}\, \mathrm{d}t)^{\frac{1}{p}}<\infty$ and $g\in\mathrm{L}_{\infty}$, if $\mathrm{ess} \sup_{t\geq0}\|g(t)\|<\infty$, where $\mathrm{ess} \sup$ denotes the essential supremum.
\subsection{Graph theory}\label{sec2.2}
Consider a multi-agent system consisting of $n$ agents, where
the communication graph is modeled by $\mathcal{G}=(\mathcal{V}, \mathcal{E})$ with the sets of nodes $\mathcal{V}=\{1, 2, \ldots, n\}$
and edges $\mathcal{E}\subseteq\mathcal{V}\times\mathcal{V}$.
For the communication graph $\mathcal{G}$, the adjacency matrix is
denoted as $\mathcal{A}=[a_{ij}]\in\mathbb{R}^{n\times n}$, $i, j=1, 2, \ldots, n$, where $a_{ii}=0$,
$a_{ij}=1$, if $(j, i)\in\mathcal{E}$, and $a_{ij}=0$, otherwise.
The Laplacian
matrix is denoted as $\mathcal{L}=[l_{ij}]\in\mathbb{R}^{n\times n}$, where $l_{ij}=-a_{ij}$, for $i\neq j$, and
$l_{ii}=\sum\nolimits_{j=1, j\neq i}^{n}a_{ij}$, $i, j=1, 2, \ldots, n$.  The communication graph is
undirected and connected if $(i, j)\in\mathcal{E}$ is equivalent to $(j, i)\in\mathcal{E}$ and there exists a path between every pair of distinct nodes.

In our work, the communication graph among the $n$ agents satisfies the following standard assumption:
\begin{assumption}\label{as1}
The communication graph is undirected and connected.
\end{assumption}
\subsection{Reference signals}\label{sec2.3}
Denote $r_{i}(t)$ as the time-varying reference signal assigned to agent $i$, $i=1, \ldots, n$, which is continuously differentiable. The average reference signal is expressed by $\bar{r}(t)=\frac{1}{n}\sum\nolimits_{i=1}^{n}r_{i}(t)$.
For simplicity, we consider that $r_{i}(t)\in\mathbb{R}$ is a scalar, but all results in our work are extendable to arbitrary dimensions.

An assumption about $r_{i}(t)$ and $\dot{r}_{i}(t)$ is introduced in the following:
\begin{assumption}\label{as2}
There exist positive constants $\epsilon_{1}$ and $\epsilon_{2}$ such that
\begin{align}
\sup_{i=1, \ldots, n}|r_{i}(t)|\leq\epsilon_{1}, \sup_{i=1, \ldots, n}|\dot{r}_{i}(t)|\leq\epsilon_{2}.\notag
\end{align}
\end{assumption}
\subsection{Problem formulation}\label{sec2.4}
The objective of this paper is to present a novel robust adaptive distributed event-based scheme using only local information to solve the DAC problem, which not only guarantees a zero estimation error, but also ensures that
the information interaction between neighboring agents is intermittent, asynchronous, and has a positive time interval.
\section{Main results}\label{sec3}
In this section, some theoretical results will be developed and some necessary discussions will be provided.
\subsection{Robust adaptive event-based DAC algorithm}\label{sec3.1}
Let $x_{i}(t)$ be the estimate of $\bar{r}(t)$. The general framework of the DAC algorithm is proposed as follows \cite{George2018cdcdistributed}:
\begin{align}
x_{i}(t)=&z_{i}(t)+r_{i}(t),\label{eq1}\\
\dot{z}_{i}(t)=&-\gamma z_{i}(t)+u_{i}(t),\label{eq2}
\end{align}
where the constant $\gamma>0$, and the input $u_{i}(t)$ will be developed subsequently.

Denote $t_{i}^{k_{i}}$, $k_{i}=0, 1, 2, \ldots$, with $t_{i}^{0}=0$, as the triggering event sequence of agent $i$, $i=1, \ldots, n$. Let $\hat{x}_{i}(t)=x_{i}(t_{i}^{k_{i}})$ for $t\in[t_{i}^{k_{i}}, t_{i}^{k_{i}+1})$, where $x_{i}(t_{i}^{k_{i}})$ denotes the latest
information that agent $i$ broadcasts to its neighbors.
Under event-based communication,
we propose the following
robust adaptive distributed input
\begin{align}\label{eq3}
u_{i}(t)=-\sum\limits_{j=1}^{n}a_{ij}c_{ij}(t)\frac{(\hat{x}_{i}(t)-\hat{x}_{j}(t))}{|\hat{x}_{i}(t)-\hat{x}_{j}(t)|+\mu(t)},
\end{align}
where $\mu(t)=\mu_{1}e^{-\mu_{2}t}$, with constants $\mu_{1}, \mu_{2}>0$, and $c_{ij}(t)$ represents the adaptive gain whose update law will be presented later.
Moreover, under the robust adaptive distributed input $u_{i}(t)$ proposed in (\ref{eq3}), the initial value of the internal state $z_{i}(t)$ is allowed to be an arbitrarily chosen bounded constant, without any constraint conditions as made in \cite{Freeman2006cdcstability, Chenfei2012tacdistributed, Shahram2012autodynamic, Kia2015ijrncdynamic, Chenfei2015tacdistributed, Moradian2017accdynamic, Zhao2017autodistributed}.
\subsection{Novel dynamic event-triggered mechanism}\label{sec3.2}
Before designing the triggering function, the measurement error is given by
\begin{align}\label{eq4}
e_{i}(t)=\hat{x}_{i}(t)-x_{i}(t).
\end{align}

For $t\in[t_{i}^{k_{i}}, t_{i}^{k_{i}+1})$, the novel dynamic triggering function $f_{i}(t)$ is updated by
\begin{align}\label{eq5}
\dot{f}_{i}(t)=\min\{\eta_{i}(t), 0\}-\delta_{i}, f_{i}(t_{i}^{k_{i}+})=\bar{f}_{i},
\end{align}
where the constants $\delta_{i}>0, \bar{f}_{i}>0$, and the internal triggering variable $\eta_{i}(t)$ is designed as
\begin{align}\label{eq6}
\eta_{i}(t)=&\alpha_{i}\sum\limits_{j=1}^{n}a_{ij}
\frac{|\hat{x}_{i}(t)-\hat{x}_{j}(t)|^{2}}{|\hat{x}_{i}(t)-\hat{x}_{j}(t)|+\mu(t)}
\notag\\
&-\beta_{i}\sum\limits_{j=1}^{n}a_{ij}c_{ij}(t)|e_{i}(t)|,
\end{align}
where
the constants $0<\alpha_{i}\leq1$ and $\beta_{i}\geq1$, and the update law of $c_{ij}(t)$ will be developed subsequently.

The triggering condition is made as follows:
\begin{align}\label{eq7}
t_{i}^{k_{i}+1}=\inf\left\{t>t_{i}^{k_{i}}\mid f_{i}(t)\leq0\right\}.
\end{align}
Once an event is triggered, $e_{i}(t)$ resets to $0$ and $f_{i}(t)$ resets to $\bar{f}_{i}$. Therefore, it always holds that $0\leq f_{i}(t)\leq\bar{f}_{i}$ and $\dot{f}_{i}(t)\leq0$.
\subsection{Composite adaptive update law}\label{sec3.3}
Partly motivated by \cite{Yucelen2013taclow, Mei2016tacdistributed}, the composite adaptive update law of $c_{ij}(t)$ is designed in the following:
\begin{align}\label{eq8}
\dot{c}_{ij}(t)=a_{ij}\frac{|\hat{x}_{i}(t)-\hat{x}_{j}(t)|^{2}}{|\hat{x}_{i}(t)-\hat{x}_{j}(t)|+\mu(t)}-\sigma_{ij}(c_{ij}(t)-\hat{c}_{ij}(t)),
\end{align}
where the constant $\sigma_{ij}=\sigma_{ji}>0$, $a_{ij}\frac{|\hat{x}_{i}(t)-\hat{x}_{j}(t)|^{2}}{|\hat{x}_{i}(t)-\hat{x}_{j}(t)|+\mu(t)}$ represents the direct adaptive component and $-\sigma_{ij}(c_{ij}(t)-\hat{c}_{ij}(t))$ represents the $\sigma$-modification component with $\hat{c}_{ij}(t)$ being the compensation term updated by
\begin{align}\label{eq9}
\dot{\hat{c}}_{ij}(t)=\nu_{ij}(c_{ij}(t)-\hat{c}_{ij}(t)),
\end{align}
where the constant $\nu_{ij}=\nu_{ji}>0$.
The initial values of $c_{ij}(t)$ and $\hat{c}_{ij}(t)$ satisfy that $c_{ij}(0)=c_{ji}(0)\geq1$, $\hat{c}_{ij}(0)=\hat{c}_{ji}(0)\geq1$, and $c_{ij}(0)\geq\hat{c}_{ij}(0)$.
\subsection{Main theorems}\label{sec3.4}
Two theorems of our work are developed in the following:
\begin{theorem}\label{thm1}
Suppose that Assumptions~\ref{as1} and \ref{as2} hold. Under the robust adaptive distributed event-based DAC algorithm (\ref{eq1})--(\ref{eq3}), the novel dynamic event-triggered mechanism (\ref{eq4})--(\ref{eq7}) and the composite adaptive update law (\ref{eq8})--(\ref{eq9}), the estimation error
\begin{align}\label{eq10}
\tilde{x}_{i}(t)=x_{i}(t)-\bar{r}(t),
\end{align}
converges to zero as time tends to infinity. Moreover, the adaptive gain $c_{ij}(t)$ and the compensation term $\hat{c}_{ij}(t)$ eventually converge to the same positive constant.
\end{theorem}
\begin{theorem}\label{thm2}
Under the same conditions stated in Theorem~\ref{thm1}, there exists a positive and uniform lower bound of the inter-event intervals for any time.
\end{theorem}

The proofs of Theorems~\ref{thm1} and \ref{thm2} will be provided in Appendices~\ref{app1} and \ref{app2}, respectively.
\subsection{Some discussions}\label{sec3.5}
To clearly show the advantages of the theoretical design in Sections~\ref{sec3.1}--\ref{sec3.3}, some comparisons with other relevant results are performed in Table~\ref{tab1}.
In addition, some properties of the proposed theoretical design are elaborated in this subsection.
\begin{table*}[!t]
  \centering
  \caption{Comparisons with some existing distributed event-based solutions to the DAC problem.}
  \label{tab1}
  \tabcolsep=0.2cm
  \renewcommand\arraystretch{1.2}
  {
  \begin{tabular}{c c c c c c c c c c}
  \hline
  reference&\cite{Kia2014cdcdynamic}&\cite{Kia2015autodistributed}&\cite{George2018cdcdistributed}&\cite{Yu2021itnecdistributed}&\cite{Xing2020tcnsrobust} &this paper\\
   \hline
  avoiding specific initialization& & &$\checkmark$&$\checkmark$&$\checkmark$&$\checkmark$ \\
  \hline
  requiring no global information&$\checkmark$&$\checkmark$& & &$\checkmark$&$\checkmark$ \\
  \hline
  guaranteeing a zero estimation error& & &$\checkmark$&$\checkmark$& &$\checkmark$\\
  \hline
  ensuring a positive and uniform minimum inter-event
interval&$\checkmark$&$\checkmark$& & &$\checkmark$&$\checkmark$\\
  \hline
  \end{tabular}}
\end{table*}
\subsubsection{Discussion on the event-based communication}\label{sec3.5.1}
\begin{figure}[!t]
\centering
{\includegraphics[height=1.5in]{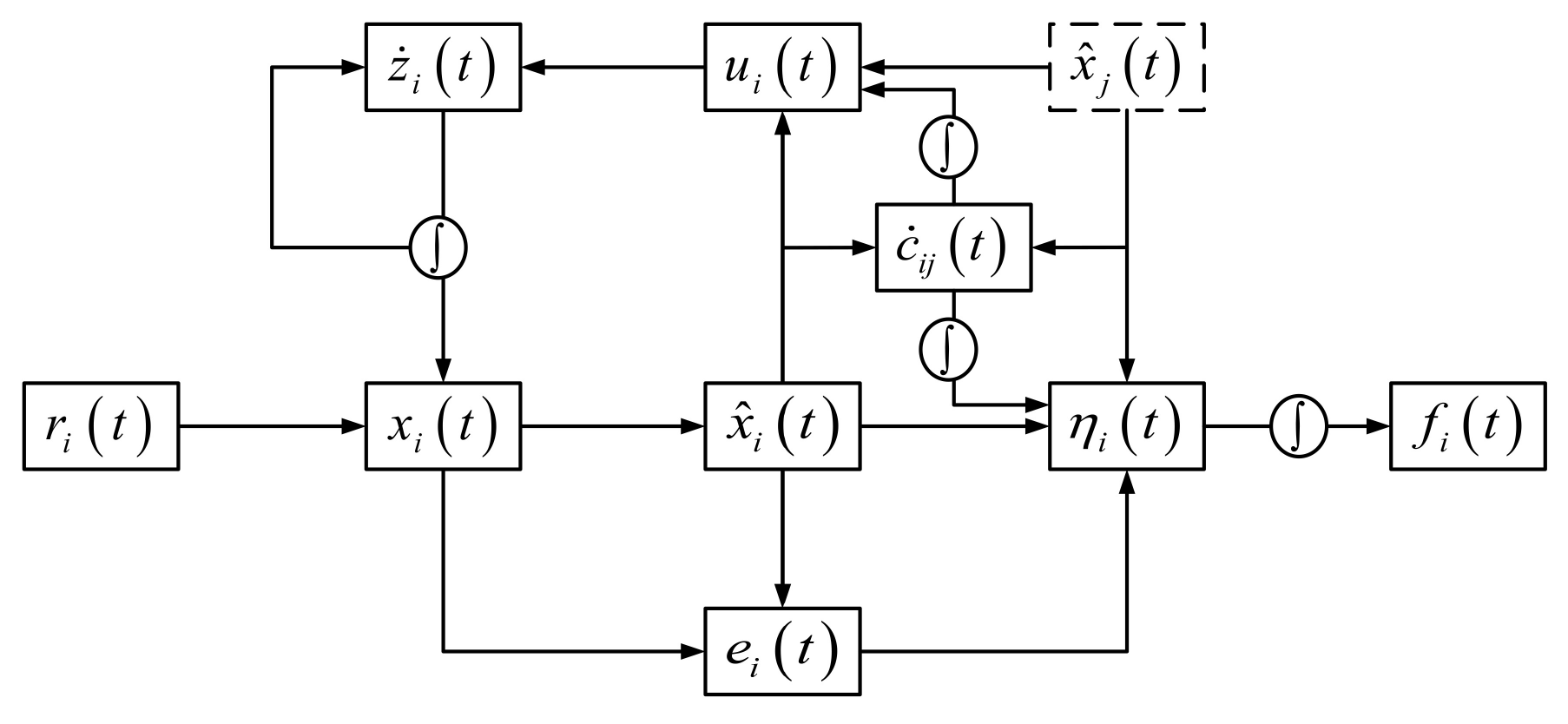}}\hspace{5pt}
\caption{A block diagram of the distributed event-based DAC solution for agent $i$ proposed in our work, where the information in dotted line box is that broadcast by neighbors of agent $i$.} \label{fig1}
\end{figure}

A block diagram of the distributed event-based DAC solution for agent $i$ proposed in Sections~\ref{sec3.1}--\ref{sec3.3} is presented in Fig.~\ref{fig1}.
It can be observed from Fig.~\ref{fig1} that the event-based intermittent state
information $\hat{x}_{i}(t)$ and $\hat{x}_{j}(t)$
is employed to design the input $u_{i}(t)$ of DAC algorithm, the internal triggering variable $\eta_{i}(t)$, and the composite adaptive update law $\dot{c}_{ij}(t)$.
Recall that $\hat{x}_{i}(t)$ and $\hat{x}_{j}(t)$ are updated only at certain instants determined by discrete events.
Therefore, the information interaction and computation of the proposed distributed event-based DAC
solution occurs only when agent $i$ or one of its neighbors trigger an event.
Compared with some distributed DAC solutions composed of continuous-time relative state $x_{i}(t)-x_{j}(t)$, which
requires continuous-time information exchange and computation,
more communication and computing resources are saved in our work.
\subsubsection{Discussion on the dynamic triggering}\label{sec3.5.2}
The dynamic property of the triggering function $f_{i}(t)$ proposed in our work lies in that $f_{i}(t)$ itself is dynamically updated, which is different from the traditional dynamic triggering functions as designed in \cite{George2018cdcdistributed, Yu2021itnecdistributed}, where only embedded internal variables are dynamically updated. Moreover, compared with the event-triggered mechanisms developed in \cite{George2018cdcdistributed, Yu2021itnecdistributed}, the implementation of the event-triggered mechanism in our work does not depend on any global information.
From (\ref{eq33}), the inter-event time of agent $i$ is lower bounded by a positive constant $\frac{\bar{f}_{i}}{\bar{\eta}_{i}+\delta_{i}}$, with $\bar{\eta}_{i}$ defined below (\ref{eq29}).
If we choose a larger $\bar{f}_{i}$ or a smaller $\delta_{i}$, the lower bound of the inter-event time will become larger.
Nevertheless, a larger $\bar{f}_{i}$ will increase $V(0)$, the initial value of $V(t)$ proposed in (\ref{eq12}),
and a smaller $\delta_{i}$ will reduce the
descent speed of $V(t)$. Therefore, under a larger $\bar{f}_{i}$ and a smaller $\delta_{i}$, the convergence rate of the estimation
error will be lowered down. Accordingly, there exists a trade-off between estimation rate and triggering performance.
\section{Simulations}\label{sec4}
The estimation effect stated in Theorem~\ref{thm1} and the triggering performance argued in Theorem~\ref{thm2} are demonstrated in this section, where the discretization period is $1\times10^{-4}\mathrm{s}$.
\subsection{Setting}\label{sec4.1}
Consider a multi-agent system involving $8$ agents, which are divided into two subgroups labeled as $\mathcal{S}_{1}=\{1, \ldots, 4\}$ and $\mathcal{S}_{2}=\{5, \ldots, 8\}$.

\begin{figure}[!t]
\centering
\subfigure[$0\mathrm{s}\leq t<6\mathrm{s}$.]{
\includegraphics[height=0.6in]{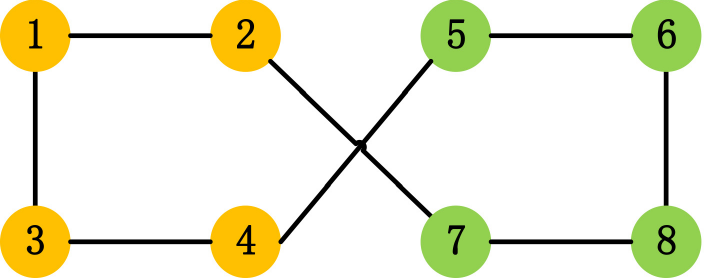}
}
\quad
\subfigure[$t\geq6\mathrm{s}$.]{
\includegraphics[height=0.6in]{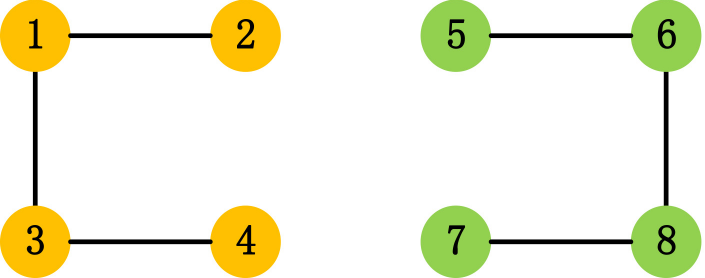}
}\hspace{5pt}
\caption{Communication graphs with network link failure.}\label{fig2}
\end{figure}

To show that the developed DAC algorithm is applicable to the case of communication disruptions,
the communication graphs with network interruptions are shown in Fig.~\ref{fig2}, where the network links between agents $2$ and $7$ and between agents $4$ and $5$ only work during the first $6\mathrm{s}$, but fail after $6\mathrm{s}$.

\begin{figure}[!t]
\centering
{\includegraphics[height=1.4in]{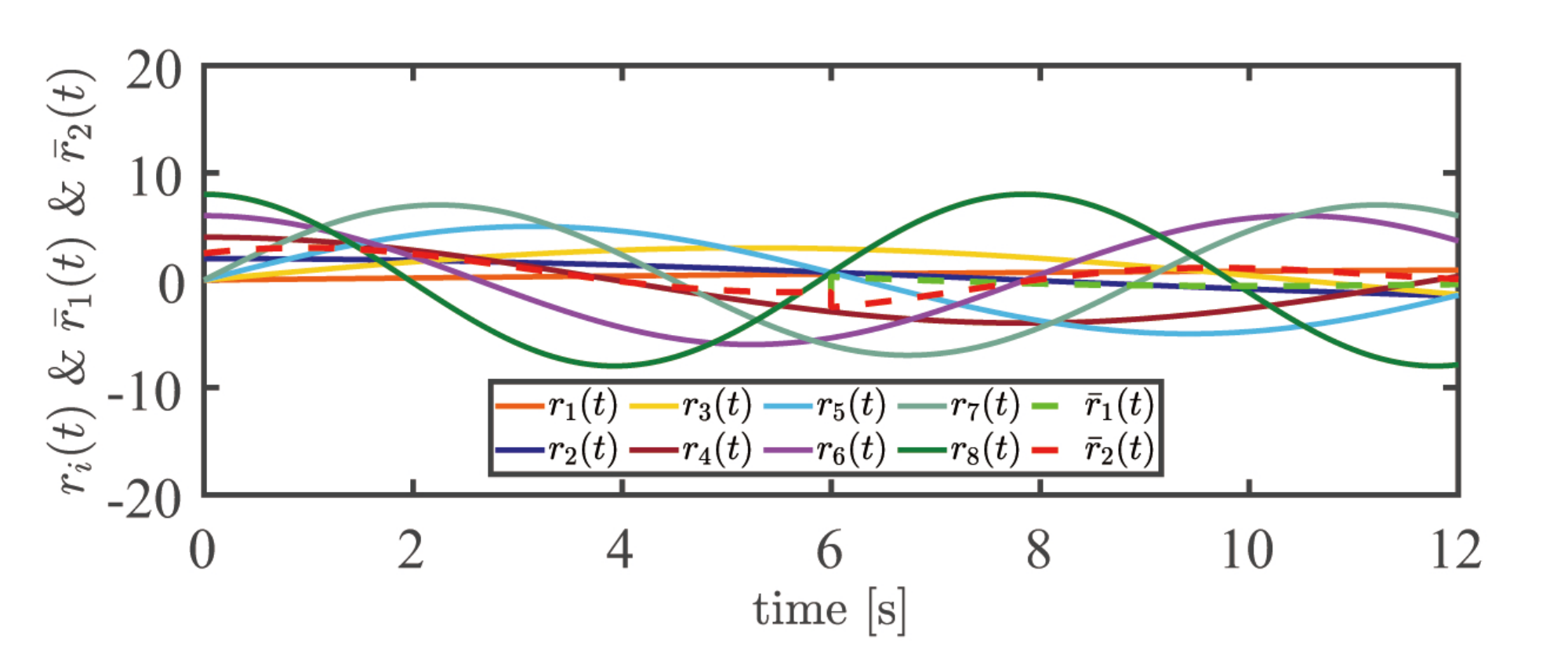}}\hspace{5pt}
\caption{The trajectories of $r_{i}(t)$, $\bar{r}_{1}(t)$ and $\bar{r}_{2}(t)$, $i=1, \ldots, 8$.} \label{fig3}
\end{figure}

For agents $1$--$8$, the individual reference signals are set as
\begin{align}
r_{i}(t)=&a_{i}\sin(b_{i}t), i=1, 3, 5, 7, \notag\\
r_{i}(t)=&a_{i}\cos(b_{i}t), i=2, 4, 6, 8,\notag
\end{align}
where $a_{i}=i$, $b_{i}=0.1i$, $i=1, \ldots, 8$.
The trajectories of the reference signals $r_{i}(t)$, $i=1, \ldots, 8$, and the average reference signals
$\bar{r}_{1}(t)=\frac{1}{8}\sum\nolimits_{i=1}^{8}r_{i}(t) (0\mathrm{s}\leq t<6\mathrm{s})
\cup\frac{1}{4}\sum\nolimits_{i=1}^{4}r_{i}(t) (t\geq6\mathrm{s})$ and
$\bar{r}_{2}(t)=\frac{1}{8}\sum\nolimits_{i=1}^{8}r_{i}(t) (0\mathrm{s}\leq t<6\mathrm{s})
\cup\frac{1}{4}\sum\nolimits_{i=5}^{8}r_{i}(t) (t\geq6\mathrm{s})$ are shown in Fig.~\ref{fig3}.

Before validating the efficiency of Theorems~\ref{thm1} and~\ref{thm2}, some initial values and designed parameters are set here. The initial values $z_{i}(0)$ are randomly chosen, $i=1, \ldots, 8$.
The initial values $c_{ij}(0)=c_{ji}(0)$ are randomly chosen from the range $[60, 80]$, and
$\hat{c}_{ij}(0)=\hat{c}_{ji}(0)=50$, $i, j=1, \ldots, 8$.
The constant gains are chosen as $\gamma=1$, $\mu_{1}=1$, $\mu_{2}=0.01$, $\bar{f}_{i}=10$, $\delta_{i}=1$, $\alpha_{i}=0.01$, $\beta_{i}=100$, $\sigma_{ij}=\sigma_{ji}=5$, $\nu_{ij}=\nu_{ji}=5$, $i, j=1, \ldots, 8$.
\subsection{Demonstration of the estimation effect}\label{sec4.2}
This subsection is to show the convergence results of the estimation error and the adaptive gain to demonstrate the effectiveness of Theorem~\ref{thm1}.

\begin{figure}[!t]
\centering
{\includegraphics[height=1.4in]{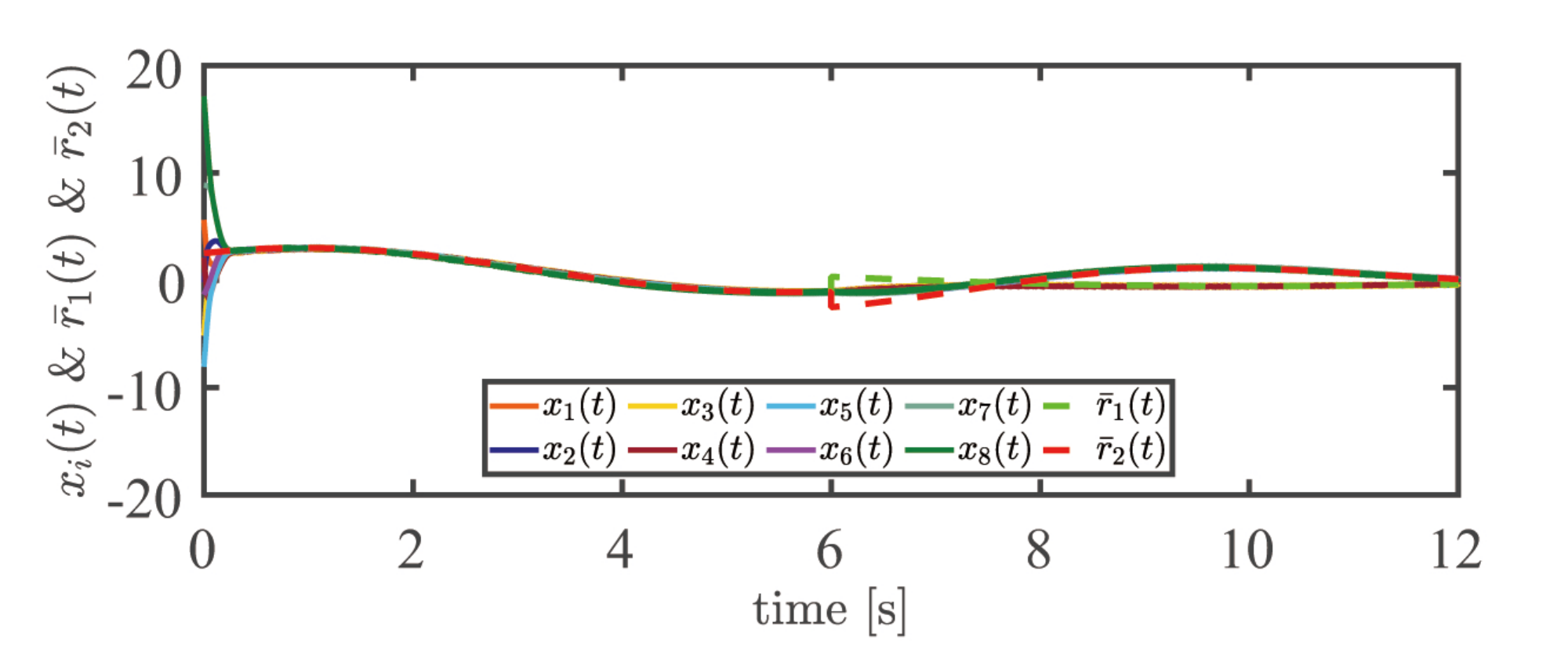}}\hspace{5pt}
\caption{The trajectories of $x_{i}(t)$, $\bar{r}_{1}(t)$ and $\bar{r}_{2}(t)$, $i=1, \ldots, 8$.} \label{fig4}
\end{figure}

\begin{figure}[!t]
\centering
{\includegraphics[height=1.4in]{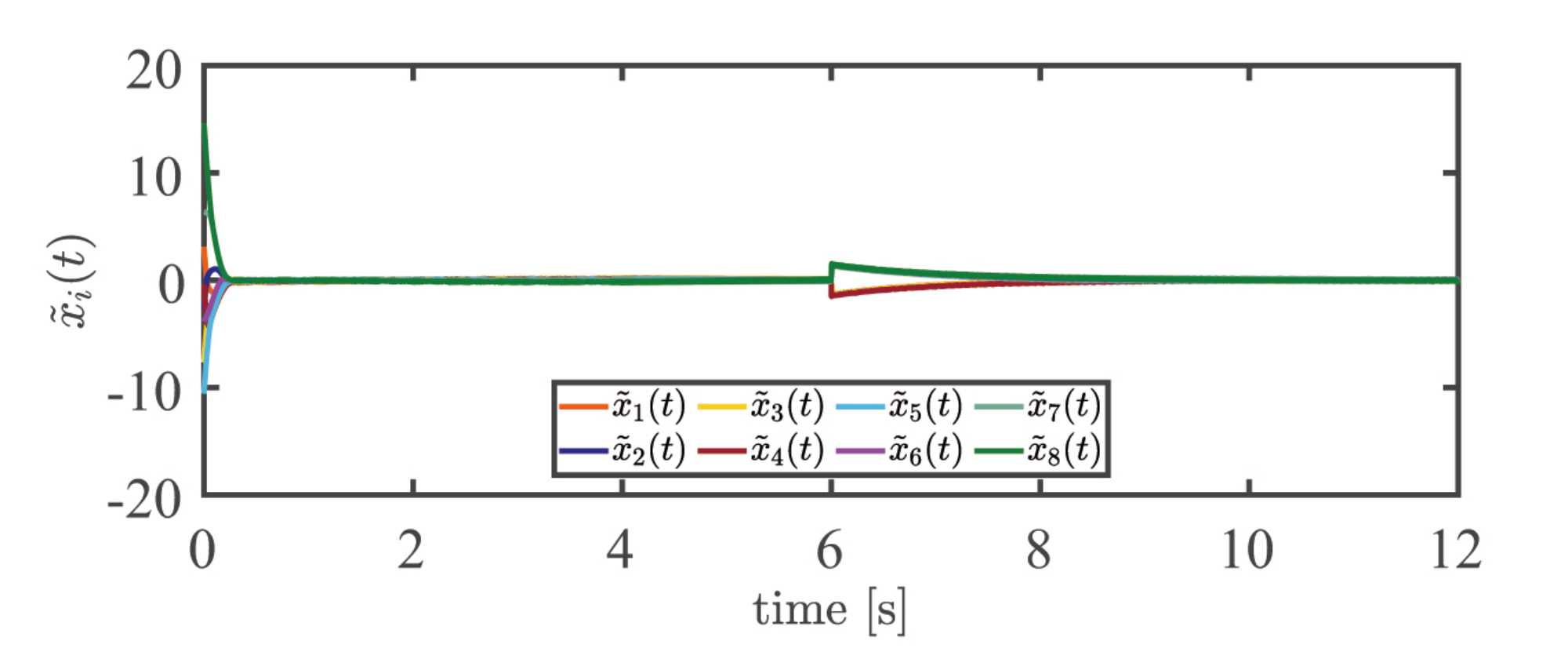}}\hspace{5pt}
\caption{The evolution of $\tilde{x}_{i}(t)$, $i=1, \ldots, 8$.} \label{fig5}
\end{figure}

\begin{figure}[!t]
\centering
{\includegraphics[height=1.4in]{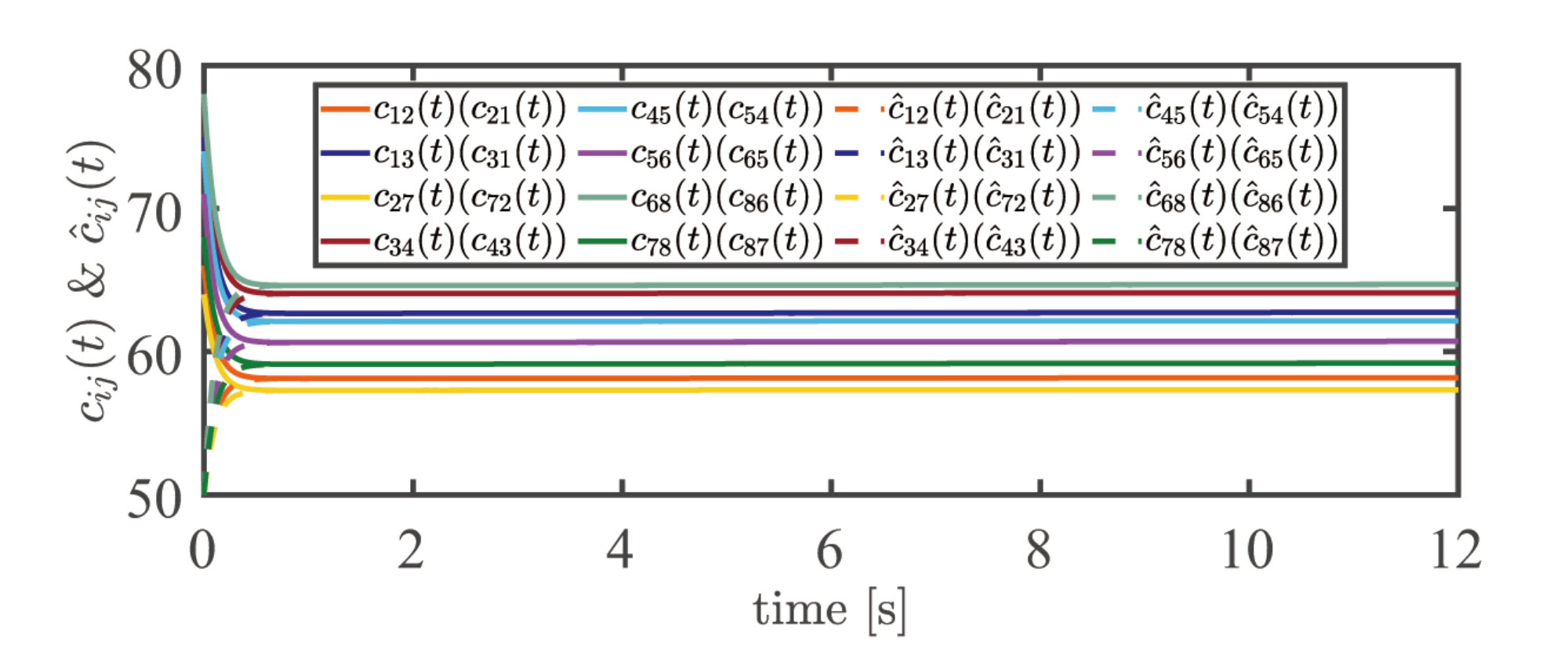}}\hspace{5pt}
\caption{The trajectories of $c_{ij}(t)$ and $\hat{c}_{ij}(t)$, $i, j=1, \ldots, 8$.} \label{fig6}
\end{figure}

Simulation contents: The trajectories of $x_{i}(t)$, $i=1, \ldots, 8$, and two average reference signals $\bar{r}_{1}(t)$ and $\bar{r}_{2}(t)$ are exhibited in Fig.~\ref{fig4}.
The evolution of the estimation errors $\tilde{x}_{i}(t)$, $i=1, \ldots, 8$, is shown in Fig.~\ref{fig5}.
The trajectories of the adaptive gains $c_{ij}(t)$ and the compensation terms $\hat{c}_{ij}(t)$, $i, j=1, \ldots, 8$, are shown in Fig.~\ref{fig6}.


Simulation observations: It follows from Fig.~\ref{fig4} that each agent achieves
accurate estimation of the average reference signals.
Fig.~\ref{fig5} shows that the estimation error of each agent converges to zero.
It can be observed from Fig.~\ref{fig6} that each adaptive gain and the corresponding compensation term converge to the same positive constant.
\subsection{Illustration of the triggering performance}\label{sec4.3}
This subsection aims to show the validity of Theorem~\ref{thm2} by providing some statistics relevant to the proposed novel dynamic event-triggered mechanism.

\begin{figure}[!t]
\centering
{\includegraphics[height=1.4in]{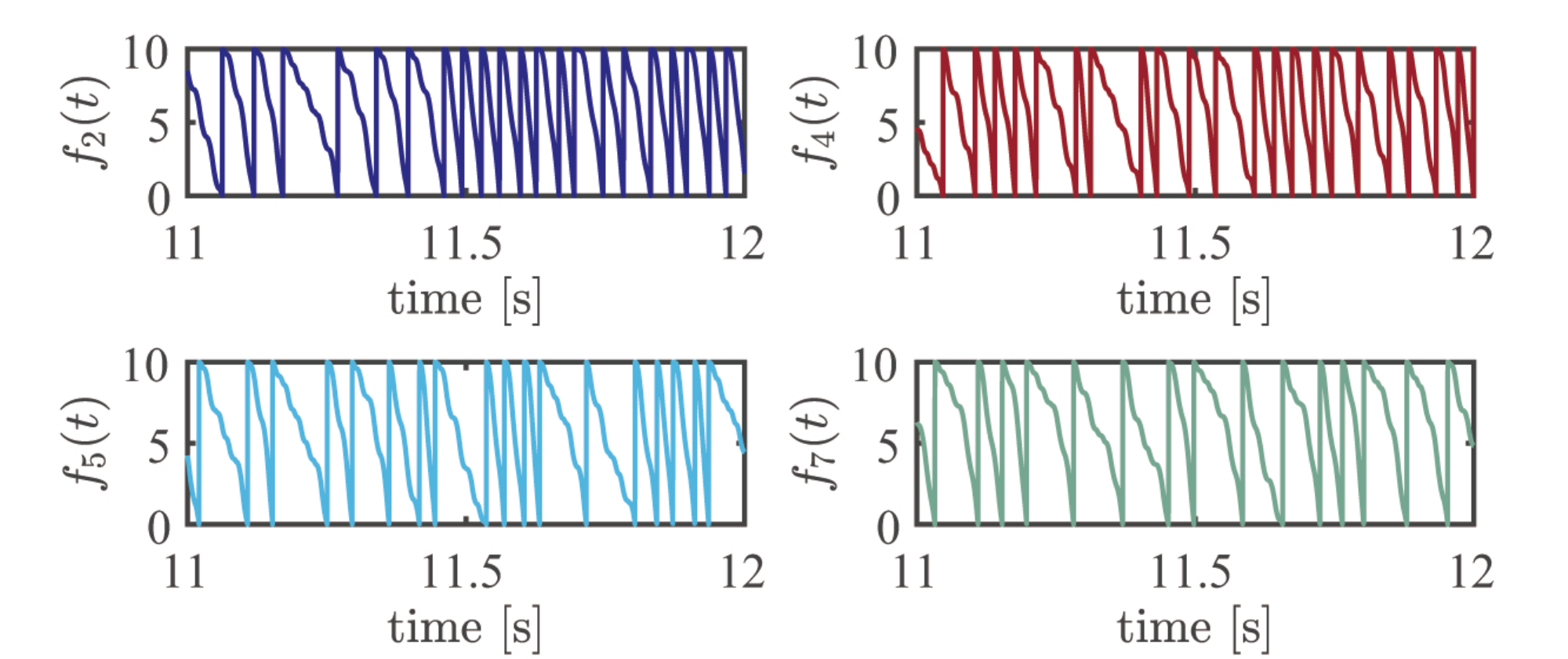}}\hspace{5pt}
\caption{The evolution of $f_{i}(t)$, $i=2, 4, 5, 7$.} \label{fig7}
\end{figure}

Simulation contents: Take agents $2, 4, 5, 7$ for example, the evolution of the dynamic triggering functions $f_{i}(t)$, $i=2, 4, 5, 7$, is shown in Fig.~\ref{fig7},
where only the trajectories from $11\mathrm{s}$ to $12\mathrm{s}$ are provided for clearer observation.
Let $\mathrm{low}_{i}=\frac{\bar{f}_{i}}{\bar{\eta}_{i}+\delta_{i}}$, which is the theoretical lower bound of inter-event time of agent $i$ derived from (\ref{eq33}).
Moreover, for each agent, we denote by $\mathrm{min}_{i}$ and $\mathrm{total}_{i}$
the minimum inter-event interval and the total number of the event instants derived from the simulation, respectively.
The numerical statistics of $\mathrm{low}_{i} [\times 10^{-4}\mathrm{s}]$, $\mathrm{min}_{i} [\times 10^{-4}\mathrm{s}]$ and $\mathrm{total}_{i}$ are presented in Table~\ref{tab2}.

\begin{table}[!t]
  \centering
  \caption{Numerical statistics of $\mathrm{low}_{i}$, $\mathrm{min}_{i}$ and $\mathrm{total}_{i}$, $i=1, \ldots, 8$.}
  \label{tab2}
  \tabcolsep=0.2cm
  \renewcommand\arraystretch{1.2}
  {
  \begin{tabular}{c c c c c c c c c}
    \hline
   &agent $1$&agent $2$&agent $3$&agent $4$ \\
  \hline
    $\mathrm{low}_{i}$&16&36&15&31\\
  \hline
    $\mathrm{min}_{i}$&35&38&32&33\\
  \hline
    $\mathrm{total}_{i}$&950&566&986&609
\\
    \hline
    &agent $5$&agent $6$&agent $7$&agent $8$ \\
  \hline
    $\mathrm{low}_{i}$&32&32&48&15\\
  \hline
    $\mathrm{min}_{i}$&33&71&59&32\\
  \hline
    $\mathrm{total}_{i}$&540&912&526&946
\\
    \hline
  \end{tabular}}
\end{table}

Simulation observations:
It follows from Fig.~\ref{fig7} that the dynamic triggering function $f_{i}(t)$ decays from $\bar{f}_{i}$ and instantly resets to $\bar{f}_{i}$ when it reaches zero.
From Table~\ref{tab2}, the minimum inter-event interval of each agent derived from the simulation is positive and larger than the theoretical lower bound of inter-event time. Moreover, it can be observed that only a finite number of events are triggered during the operation time.
\section{Conclusion}\label{sec5}
A distributed event-based solution is provided in this paper to deal with the DAC problem, where the estimation error converges to zero.
The proposed solution mainly involves three parts, namely the
distributed event-based DAC algorithm, the
distributed dynamic event-triggered mechanism, and the composite
adaptive update law of the adaptive gain.
Local relative information, which is updated only at certain discrete event instants, is employed in the solution, such that continuous information exchange among neighboring agents is avoided.
The developed distributed event-based DAC algorithm is robust in the sense that no specific initialization criterion is required during the implementation.
Under the developed dynamic event-triggered mechanism, the inter-event intervals of each agent are lower bounded by a positive constant for all time, definitely excluding the Zeno behavior. Moreover, using the adaptive gain together with its composite update law, the proposed solution is performed
 without requiring any global information.
The effect of disturbance on the distributed DAC solution design and analysis will be studied in our future work.
\appendices
\section{Proof of Theorem~\ref{thm1}}\label{app1}
Before verifying the feasibility of Theorem~\ref{thm1}, a useful lemma is introduced as follows.
\begin{lemma}\label{le1}\cite{George2019tacrobust, godsil2001algebraic}
Suppose that Assumption~\ref{as1} holds. The Laplacian matrix $\mathcal{L}$ is symmetric positive-semidefinite with $n$ eigenvalues ordered
by $0=\lambda_{1}<\lambda_{2}\leq\ldots\leq\lambda_{n}$. Moreover, for $\mathcal{M}=I_{n}-\frac{1}{n}\mathbf{1}_{n}\mathbf{1}_{n}^{T}$, it holds that
$\mathcal{M}=\mathcal{L}\mathcal{L}^{\dagger}$.
\end{lemma}

The proof of Theorem~\ref{thm1} is detailed as below.

\begin{proof}
Let $x(t)$, $\tilde{x}(t)$, $z(t)$, $r(t)$, and $u(t)$ be the column stack vectors of $x_{i}(t)$, $\tilde{x}_{i}(t)$, $z_{i}(t)$, $r_{i}(t)$, and $u_{i}(t)$, $i=1, \ldots, n$, respectively.

The Lyapunov function is constructed as
\begin{align}\label{eq12}
V(t)=V_{1}(t)+V_{2}(t)+V_{3}(t),
\end{align}
where
\begin{subequations}
\begin{align}
V_{1}(t)=&\frac{1}{2}\sum\limits_{i=1}^{n}\tilde{x}_{i}^{2}(t),\label{eq13a}\\
V_{2}(t)=&\theta_{1}\sum\limits_{i=1}^{n}f_{i}(t),\label{eq13b}\\
V_{3}(t)=&\frac{1}{4}\sum\limits_{i=1}^{n}\sum\limits_{j=1}^{n}
\left[(c_{ij}(t)-\theta_{2})^{2}+\frac{\sigma_{ij}}{\nu_{ij}}(\hat{c}_{ij}(t)-\theta_{2})^{2}\right],\label{eq13c}
\end{align}
\end{subequations}
where the constants $\theta_{1}\geq1+2\bar{\epsilon}$ and
$\theta_{2}\geq2+6\bar{\epsilon}$, with $\bar{\epsilon}=\frac{\sqrt{n}}{\lambda_{2}}(\gamma\epsilon_{1}
+\epsilon_{2})>0$.

Note that $c_{ij}(0)=c_{ji}(0)$, $\hat{c}_{ij}(0)=\hat{c}_{ji}(0)$, $\sigma_{ij}=\sigma_{ji}>0$, $\nu_{ij}=\nu_{ji}>0$.
Note also that $a_{ij}=a_{ji}$.
From (\ref{eq8}) and (\ref{eq9}), it is easy to verify that $c_{ij}(t)=c_{ji}(t)$, $\hat{c}_{ij}(t)=\hat{c}_{ji}(t)$.
It follows from (\ref{eq8}) that $c_{ij}(t)$ will instantly increase if $c_{ij}(t)\leq\hat{c}_{ij}(t)$, since $\dot{c}_{ij}(t)\geq0$ at the instant when $c_{ij}(t)\leq\hat{c}_{ij}(t)$.
At the same time, from (\ref{eq9}) $c_{ij}(t)\leq\hat{c}_{ij}(t)$ implies that
$\dot{\hat{c}}_{ij}(t)\leq0$. Hence, $\hat{c}_{ij}(t)$ will instantly decrease if $c_{ij}(t)\leq\hat{c}_{ij}(t)$.
Combining $c_{ij}(0)\geq\hat{c}_{ij}(0)$ leads to that $c_{ij}(t)\geq\hat{c}_{ij}(t)$. It further follows from (\ref{eq9}) that $\dot{\hat{c}}_{ij}(t)\geq0$, which indicates that $\hat{c}_{ij}(t)\geq1$ since $\hat{c}_{ij}(0)\geq1$. Therefore, it always holds that $c_{ij}(t)\geq\hat{c}_{ij}(t)\geq1$.

Substituting (\ref{eq1}) into (\ref{eq10}) and using vector notation, one obtains
\begin{align}\label{eq14}
\tilde{x}(t)
=z(t)+r(t)-\bar{r}(t)\mathbf{1}_{n}
=z(t)+\mathcal{M}r(t).
\end{align}
It follows from (\ref{eq2}) and (\ref{eq14}) that
\begin{align}\label{eq15}
\dot{\tilde{x}}(t)=&\dot{z}(t)+\mathcal{M}\dot{r}(t)\notag\\
=&-\gamma \left(z(t)+\mathcal{M}r(t)\right)+u(t)
+\gamma\mathcal{M}r(t)+\mathcal{M}\dot{r}(t)\notag\\
=&-\gamma\tilde{x}(t)+u(t)+\mathcal{M}\left(\gamma r(t)+\dot{r}(t)\right).
\end{align}

Taking the derivative of $V_{1}(t)$ along the trajectory of (\ref{eq15}), one derives
\begin{align}\label{eq16}
\dot{V}_{1}(t)=&-\gamma\tilde{x}^{T}(t)\tilde{x}(t)
+\tilde{x}^{T}(t)u(t)\notag\\
&+\tilde{x}^{T}(t)\mathcal{M}\left(\gamma r(t)+\dot{r}(t)\right).
\end{align}
Since $a_{ij}=a_{ji}$, $c_{ij}(t)=c_{ji}(t)$, $\beta_{i}\geq1$, and
$\frac{(\hat{x}_{i}(t)-\hat{x}_{j}(t))}{|\hat{x}_{i}(t)-\hat{x}_{j}(t)|+\mu(t)}\leq1$,
it follows that
\begin{align}\label{eq17}
&\frac{1}{2}\sum\limits_{i=1}^{n}\sum\limits_{j=1}^{n}a_{ij}c_{ij}(t)
(e_{i}(t)-e_{j}(t))\frac{(\hat{x}_{i}(t)-\hat{x}_{j}(t))}{|\hat{x}_{i}(t)-\hat{x}_{j}(t)|+\mu(t)}\notag\\
\leq&
\frac{1}{2}\sum\limits_{i=1}^{n}\sum\limits_{j=1}^{n}a_{ij}c_{ij}(t)
|e_{i}(t)-e_{j}(t)|\notag\\
\leq&
\frac{1}{2}\sum\limits_{i=1}^{n}\sum\limits_{j=1}^{n}a_{ij}c_{ij}(t)
|e_{i}(t)|+\frac{1}{2}\sum\limits_{i=1}^{n}\sum\limits_{j=1}^{n}a_{ij}c_{ij}(t)|e_{j}(t)|\notag\\
\leq&\sum\limits_{i=1}^{n}\sum\limits_{j=1}^{n}\beta_{i}a_{ij}c_{ij}(t)
|e_{i}(t)|.
\end{align}
From (\ref{eq4}) and (\ref{eq10}), one has
$\tilde{x}_{i}(t)-\tilde{x}_{j}(t)=x_{i}(t)-x_{j}(t)=
(\hat{x}_{i}(t)-\hat{x}_{j}(t))-(e_{i}(t)-e_{j}(t))$.
It follows from (\ref{eq3}) and (\ref{eq17}) that
\begin{align}\label{eq18}
&\tilde{x}^{T}(t)u(t)\notag\\
=&-\sum\limits_{i=1}^{n}\sum\limits_{j=1}^{n}a_{ij}c_{ij}(t)\tilde{x}_{i}(t)\frac{(\hat{x}_{i}(t)-\hat{x}_{j}(t))}{|\hat{x}_{i}(t)-\hat{x}_{j}(t)|+\mu(t)}\notag\\
=&-\frac{1}{2}\sum\limits_{i=1}^{n}\sum\limits_{j=1}^{n}a_{ij}c_{ij}(t)(\tilde{x}_{i}(t)-\tilde{x}_{j}(t))\frac{(\hat{x}_{i}(t)-\hat{x}_{j}(t))}{|\hat{x}_{i}(t)-\hat{x}_{j}(t)|+\mu(t)}\notag\\
=&-\frac{1}{2}\sum\limits_{i=1}^{n}\sum\limits_{j=1}^{n}a_{ij}c_{ij}(t)(\hat{x}_{i}(t)-\hat{x}_{j}(t))\frac{(\hat{x}_{i}(t)-\hat{x}_{j}(t))}{|\hat{x}_{i}(t)-\hat{x}_{j}(t)|+\mu(t)}\notag\\
&+\frac{1}{2}\sum\limits_{i=1}^{n}\sum\limits_{j=1}^{n}a_{ij}c_{ij}(t)(e_{i}(t)-e_{j}(t))\frac{(\hat{x}_{i}(t)-\hat{x}_{j}(t))}{|\hat{x}_{i}(t)-\hat{x}_{j}(t)|+\mu(t)}\notag\\
\leq&-\frac{1}{2}\sum\limits_{i=1}^{n}\sum\limits_{j=1}^{n}a_{ij}c_{ij}(t)\frac{|\hat{x}_{i}(t)-\hat{x}_{j}(t)|^{2}}{|\hat{x}_{i}(t)-\hat{x}_{j}(t)|+\mu(t)}\notag\\
&+\sum\limits_{i=1}^{n}\sum\limits_{j=1}^{n}\beta_{i}a_{ij}c_{ij}(t)|e_{i}(t)|.
\end{align}
Similarly as in deriving (\ref{eq17}), one has
\begin{align}\label{eq19}
\sum\limits_{i=1}^{n}\sum\limits_{j=1}^{n}a_{ij}|e_{i}(t)-e_{j}(t)|
\leq
2\sum\limits_{i=1}^{n}\sum\limits_{j=1}^{n}\beta_{i}a_{ij}c_{ij}(t)|e_{i}(t)|,
\end{align}
where $a_{ij}=a_{ji}$, $\beta_{i}\geq1$, and $c_{ij}(t)\geq1$ have been used.
It follows from Lemma~\ref{le1} that $\mathcal{L}=\mathcal{L}^{T}$ and $\mathcal{M}=\mathcal{L}\mathcal{L}^{\dagger}$.
Let $y(t)=\mathcal{L}^{\dagger}(\gamma r(t)+\dot{r}(t))$.
Under Assumption~\ref{as2} and Lemma~\ref{le1}, it holds that
$\|y(t)\|\leq\|\mathcal{L}^{\dagger}\|(\|\gamma r(t)\|+\|\dot{r}(t)\|)\leq\bar{\epsilon}$, where $\bar{\epsilon}$ has been defined below (\ref{eq13c}).
Based on the above analysis, one obtains
\begin{align}\label{eq20}
&\tilde{x}^{T}(t)\mathcal{M}\left(\gamma r(t)+\dot{r}(t)\right)\notag\\
=&\tilde{x}^{T}(t)\mathcal{L}y(t)=y^{T}(t)\mathcal{L}\tilde{x}(t)\notag\\
\leq&\sum\limits_{i=1}^{n}\sum\limits_{j=1}^{n}a_{ij}|y_{i}(t)||\tilde{x}_{i}(t)-\tilde{x}_{j}(t)|\notag\\
\leq&\bar{\epsilon}\sum\limits_{i=1}^{n}\sum\limits_{j=1}^{n}a_{ij}\left(|\hat{x}_{i}(t)-\hat{x}_{j}(t)|
+|e_{i}(t)-e_{j}(t)|\right)\notag\\
\leq&\bar{\epsilon}\sum\limits_{i=1}^{n}\sum\limits_{j=1}^{n}a_{ij}
|\hat{x}_{i}(t)-\hat{x}_{j}(t)|
+2\bar{\epsilon}\sum\limits_{i=1}^{n}\sum\limits_{j=1}^{n}\beta_{i}a_{ij}c_{ij}(t)|e_{i}(t)|,
\end{align}
where $y_{i}(t)$ represents the $i$-th entry of $y(t)$ satisfying that $|y_{i}(t)|\leq\|y(t)\|\leq\bar{\epsilon}$.

Combining (\ref{eq16}), (\ref{eq18}) and (\ref{eq20}) leads to that
\begin{align}\label{eq21}
\dot{V}_{1}(t)
\leq&-\gamma\sum\limits_{i=1}^{n}\tilde{x}_{i}^{2}(t)
+\left(1+2\bar{\epsilon}\right)\sum\limits_{i=1}^{n}\sum\limits_{j=1}^{n}\beta_{i}a_{ij}
c_{ij}(t)
|e_{i}(t)|\notag\\
&-\frac{1}{2}\sum\limits_{i=1}^{n}\sum\limits_{j=1}^{n}a_{ij}c_{ij}(t)\frac{|\hat{x}_{i}(t)-\hat{x}_{j}(t)|^{2}}{|\hat{x}_{i}(t)-\hat{x}_{j}(t)|+\mu(t)}\notag\\
&+\bar{\epsilon}\sum\limits_{i=1}^{n}\sum\limits_{j=1}^{n}a_{ij}|\hat{x}_{i}(t)-\hat{x}_{j}(t)|.
\end{align}

Taking the derivative of $V_{2}(t)$ along the trajectory of (\ref{eq5}), one obtains
$\dot{V}_{2}(t)=\theta_{1}\sum\nolimits_{i=1}^{n}\dot{f}_{i}(t)$. Since $\dot{f}_{i}(t)\leq0$ and $\theta_{1}\geq1+2\bar{\epsilon}$, one obtains
\begin{align}\label{eq22}
\dot{V}_{2}(t)\leq&(1+2\bar{\epsilon})\sum\limits_{i=1}^{n}\dot{f}_{i}(t)\notag\\
=&
(1+2\bar{\epsilon})\sum\limits_{i=1}^{n}\left(\min\{\eta_{i}(t), 0\}-\delta_{i}\right)\notag\\
\leq&(1+2\bar{\epsilon})\sum\limits_{i=1}^{n}\eta_{i}(t).
\end{align}

The derivative of $V_{3}(t)$ along the trajectories of (\ref{eq8}) and (\ref{eq9}) is calculated in the following:
\begin{align}\label{eq23}
\dot{V}_{3}(t)
=&\frac{1}{2}\sum\limits_{i=1}^{n}\sum\limits_{j=1}^{n}a_{ij}(c_{ij}(t)-\theta_{2})\frac{|\hat{x}_{i}(t)-\hat{x}_{j}(t)|^{2}}{|\hat{x}_{i}(t)-\hat{x}_{j}(t)|+\mu(t)}\notag\\
&-\frac{1}{2}\sum\limits_{i=1}^{n}\sum\limits_{j=1}^{n}\sigma_{ij}(c_{ij}(t)-\theta_{2})(c_{ij}(t)-\hat{c}_{ij}(t))\notag\\
&+\frac{1}{2}\sum\limits_{i=1}^{n}\sum\limits_{j=1}^{n}\sigma_{ij}(\hat{c}_{ij}(t)-\theta_{2})(c_{ij}(t)-\hat{c}_{ij}(t))\notag\\
=&\frac{1}{2}\sum\limits_{i=1}^{n}\sum\limits_{j=1}^{n}a_{ij}(c_{ij}(t)-\theta_{2})\frac{|\hat{x}_{i}(t)-\hat{x}_{j}(t)|^{2}}{|\hat{x}_{i}(t)-\hat{x}_{j}(t)|+\mu(t)}\notag\\
&-\frac{1}{2}\sum\limits_{i=1}^{n}\sum\limits_{j=1}^{n}\sigma_{ij}(c_{ij}(t)-\hat{c}_{ij}(t))^{2}\notag\\
\leq&\frac{1}{2}\sum\limits_{i=1}^{n}\sum\limits_{j=1}^{n}a_{ij}(c_{ij}(t)-\theta_{2})\frac{|\hat{x}_{i}(t)-\hat{x}_{j}(t)|^{2}}{|\hat{x}_{i}(t)-\hat{x}_{j}(t)|+\mu(t)}.
\end{align}

Combining (\ref{eq21})--(\ref{eq23}) yields that
\begin{align}\label{eq24}
\dot{V}(t)
\leq&-\gamma\sum\limits_{i=1}^{n}\tilde{x}_{i}^{2}(t)
+(1+2\bar{\epsilon})\sum\limits_{i=1}^{n}\eta_{i}(t)\notag\\
&-\frac{\theta_{2}}{2}\sum\limits_{i=1}^{n}\sum\limits_{j=1}^{n}a_{ij}\frac{|\hat{x}_{i}(t)-\hat{x}_{j}(t)|^{2}}{|\hat{x}_{i}(t)-\hat{x}_{j}(t)|+\mu(t)}\notag\\
&+\bar{\epsilon}\sum\limits_{i=1}^{n}\sum\limits_{j=1}^{n}a_{ij}\left|\hat{x}_{i}(t)-\hat{x}_{j}(t)\right|\notag\\
&+\left(1+2\bar{\epsilon}\right)\sum\limits_{i=1}^{n}
\sum\limits_{j=1}^{n}\beta_{i}a_{ij}c_{ij}(t)|e_{i}(t)|.
\end{align}
Note that
$\theta_{2}\geq2+6\bar{\epsilon}$ and $0<\alpha_{i}\leq1$. Through simple calculations, one has
\begin{align}\label{eq25}
&-\frac{\theta_{2}}{2}\sum\limits_{i=1}^{n}\sum\limits_{j=1}^{n}a_{ij}\frac{|\hat{x}_{i}(t)-\hat{x}_{j}(t)|^{2}}{|\hat{x}_{i}(t)-\hat{x}_{j}(t)|+\mu(t)}\notag\\
&+\bar{\epsilon}\sum\limits_{i=1}^{n}\sum\limits_{j=1}^{n}a_{ij}\left|\hat{x}_{i}(t)-\hat{x}_{j}(t)\right|\notag\\
\leq&-\left(1+2\bar{\epsilon}\right)\sum\limits_{i=1}^{n}\sum\limits_{j=1}^{n}a_{ij}\frac{|\hat{x}_{i}(t)-\hat{x}_{j}(t)|^{2}}{|\hat{x}_{i}(t)-\hat{x}_{j}(t)|+\mu(t)}\notag\\
&+\bar{\epsilon}\sum\limits_{i=1}^{n}\sum\limits_{j=1}^{n}a_{ij}\frac{\mu(t)|\hat{x}_{i}(t)-\hat{x}_{j}(t)|}{|\hat{x}_{i}(t)-\hat{x}_{j}(t)|+\mu(t)}\notag\\
\leq&-\left(1+2\bar{\epsilon}\right)\sum\limits_{i=1}^{n}\sum\limits_{j=1}^{n}\alpha_{i}a_{ij}\frac{|\hat{x}_{i}(t)-\hat{x}_{j}(t)|^{2}}{|\hat{x}_{i}(t)-\hat{x}_{j}(t)|+\mu(t)}+\bar{\mu}_{1}e^{-\mu_{2}t},
\end{align}
where $\frac{\mu(t)|\hat{x}_{i}(t)-\hat{x}_{j}(t)|}{|\hat{x}_{i}(t)-\hat{x}_{j}(t)|+\mu(t)}\leq\mu(t)=
\mu_{1}e^{-\mu_{2}t}$ has been used and
$\bar{\mu}_{1}=\bar{\epsilon}n^{2}\mu_{1}>0$. It follows from (\ref{eq6}), (\ref{eq24}) and (\ref{eq25}) that
\begin{align}\label{eq26}
\dot{V}(t)
\leq&-\gamma\sum\limits_{i=1}^{n}\tilde{x}_{i}^{2}(t)+\bar{\mu}_{1}e^{-\mu_{2}t}
+(1+2\bar{\epsilon})\sum\limits_{i=1}^{n}\eta_{i}(t)\notag\\
&-\left(1+2\bar{\epsilon}\right)\sum\limits_{i=1}^{n}
\sum\limits_{j=1}^{n}\alpha_{i}a_{ij}\frac{|\hat{x}_{i}(t)-\hat{x}_{j}(t)|^{2}}{|\hat{x}_{i}(t)-\hat{x}_{j}(t)|+\mu(t)}\notag\\
&+\left(1+2\bar{\epsilon}\right)\sum\limits_{i=1}^{n}
\sum\limits_{j=1}^{n}\beta_{i}a_{ij}c_{ij}(t)|e_{i}(t)|\notag\\
=&-\gamma\sum\limits_{i=1}^{n}\tilde{x}_{i}^{2}(t)+\bar{\mu}_{1}e^{-\mu_{2}t}.
\end{align}

Integrating both sides of $\dot{V}(t)
\leq-\gamma\sum\nolimits_{i=1}^{n}\tilde{x}_{i}^{2}(t)+\bar{\mu}_{1}e^{-\mu_{2}t}
=-\gamma\|\tilde{x}(t)\|^{2}+\bar{\mu}_{1}e^{-\mu_{2}t}$, one has
\begin{align}\label{eq27}
V(t)+\gamma\int_{0}^{t}\|\tilde{x}(\varsigma)\|^{2}\, \mathrm{d}\varsigma
\leq&V(0)+\frac{\bar{\mu}_{1}}{\mu_{2}}\left(1-e^{-\mu_{2}t}\right)\notag\\
\leq&V(0)+\frac{\bar{\mu}_{1}}{\mu_{2}}.
\end{align}
On the one hand, it can be derived from (\ref{eq27}) that
$V(t)\in\mathrm{L}_{\infty}$. It follows from (\ref{eq13a}) that
$\tilde{x}_{i}(t)\in\mathrm{L}_{\infty}$ and from (\ref{eq13c}) that $c_{ij}(t), \hat{c}_{ij}(t)\in\mathrm{L}_{\infty}$.
Under Assumption~\ref{as2}, $r_{i}(t), \dot{r}_{i}(t)\in\mathrm{L}_{\infty}$, and hence $\bar{r}(t)\in\mathrm{L}_{\infty}$.
It follows form (\ref{eq10}) that $x_{i}(t)\in\mathrm{L}_{\infty}$, which indicates that $\hat{x}_{i}(t)\in\mathrm{L}_{\infty}$.
From (\ref{eq3}), one derives that
$u_{i}(t)\in\mathrm{L}_{\infty}$. It thus follows from (\ref{eq15}) that $\dot{\tilde{x}}_{i}(t)\in\mathrm{L}_{\infty}$.
On the other hand, it can be derived from (\ref{eq27}) that $\tilde{x}_{i}(t)\in\mathrm{L}_{2}$.
By Barbalat's lemma \cite[lemma 3.2.5]{Ioannou1996robust}, one concludes that $\tilde{x}_{i}(t)\rightarrow0$, as $t\rightarrow\infty$.

Let $\tilde{c}_{ij}(t)=c_{ij}(t)-\hat{c}_{ij}(t)$. It follows from (\ref{eq8}) and (\ref{eq9}) that
\begin{align}\label{eq28}
\dot{\tilde{c}}_{ij}(t)=-(\sigma_{ij}+\nu_{ij})\tilde{c}_{ij}(t)
+a_{ij}\frac{|\hat{x}_{i}(t)-\hat{x}_{j}(t)|^{2}}{|\hat{x}_{i}(t)-\hat{x}_{j}(t)|+\mu(t)}.
\end{align}
Note that $\frac{|\hat{x}_{i}(t)-\hat{x}_{j}(t)|^{2}}{|\hat{x}_{i}(t)-\hat{x}_{j}(t)|+\mu(t)}\in\mathrm{L}_{\infty}$ since $\hat{x}_{i}(t)\in\mathrm{L}_{\infty}$.
Note also that
$x_{i}(t)-x_{j}(t)=\tilde{x}_{i}(t)-\tilde{x}_{j}(t)\rightarrow0$ as $t\rightarrow\infty$ since $\tilde{x}_{i}(t)\rightarrow0$, as $t\rightarrow\infty$, which implies that
$\frac{|\hat{x}_{i}(t)-\hat{x}_{j}(t)|^{2}}{|\hat{x}_{i}(t)-\hat{x}_{j}(t)|+\mu(t)}
\rightarrow0$ as $t\rightarrow\infty$.
By \cite[Theorem 4]{Sunzhiyong2020cyberevent}, one concludes that $\tilde{c}_{ij}(t)\rightarrow0$ as $t\rightarrow\infty$, i.e., $c_{ij}(t)\rightarrow\hat{c}_{ij}(t)$ as $t\rightarrow\infty$.
In addition, note that $\hat{c}_{ij}(t)$ remains bounded because $\hat{c}_{ij}(t)\in\mathrm{L}_{\infty}$. Combining $\dot{\hat{c}}_{ij}(t)\geq0$ yields that $\hat{c}_{ij}(t)$ converges to a positive constant.
Therefore, $c_{ij}(t)$ and $\hat{c}_{ij}(t)$ eventually converge to the same positive constant.
\end{proof}
\section{Proof of Theorem~\ref{thm2}}\label{app2}
The proof of Theorem~\ref{thm2} is detailed as below.

\begin{proof}
It follows from (\ref{eq6}) that
\begin{align}\label{eq29}
\eta_{i}(t)\geq-\beta_{i}\sum\limits_{j=1}^{n}a_{ij}
c_{ij}(t)|e_{i}(t)|.
\end{align}
From the proof of Theorem~\ref{thm1}, one has that $x_{i}(t)$ and $c_{ij}(t)$ remain bounded. It follows from (\ref{eq4}) that
$e_{i}(t)$ is bounded.
Without loss of generality, we assume that $c_{ij}(t)\leq\bar{c}_{ij}$, $|e_{i}(t)|\leq\bar{e}_{i}$, where the constants
$\bar{c}_{ij}, \bar{e}_{i}>0$.
It thus follows from (\ref{eq29}) that
$\eta_{i}(t)\geq-\bar{\eta}_{i}$, where the constant $\bar{\eta}_{i}=
\beta_{i}\sum\nolimits_{j=1}^{n}a_{ij}\bar{c}_{ij}
\bar{e}_{i}>0$. From (\ref{eq5}), one has
\begin{align}\label{eq30}
\dot{f}_{i}(t)\geq-(\bar{\eta}_{i}+\delta_{i}), f_{i}(t_{i}^{k_{i}+})=\bar{f}_{i}.
\end{align}
By the comparison lemma \cite[Lemma 2.5]{Khalil2002}, it holds that $f_{i}(t)\geq g_{i}(t)$ for $t\in[t_{i}^{k_{i}}, t_{i}^{k_{i}+1})$, with $g_{i}(t)$ being the solution of
\begin{align}\label{eq31}
\dot{g}_{i}(t)=-(\bar{\eta}_{i}+\delta_{i}), g_{i}(t_{i}^{k_{i}+})=\bar{f}_{i}.
\end{align}
Solving the equation (\ref{eq31}) obtains that
\begin{align}\label{eq32}
t-t_{i}^{k_{i}}=\frac{\bar{f}_{i}}{\bar{\eta}_{i}+\delta_{i}}-\frac{g_{i}(t)}{\bar{\eta}_{i}+\delta_{i}}.
\end{align}
Note that $f_{i}(t)\geq g_{i}(t)$.
Therefore, the inter-event time $t_{i}^{k_{i}+1}-t_{i}^{k_{i}}$ is lower bounded by the time that it takes for $g_{i}(t)$ to decrease from $\bar{f}_{i}$ to $0$. It thus follows from (\ref{eq32}) that
\begin{align}\label{eq33}
t_{i}^{k_{i}+1}-t_{i}^{k_{i}}\geq\frac{\bar{f}_{i}}{\bar{\eta}_{i}+\delta_{i}}.
\end{align}
It is concluded from (\ref{eq33}) that the lower bound of the inter-event time $t_{i}^{k_{i}+1}-t_{i}^{k_{i}}$ is positive for any time $t$ and uniform for any index $k_{i}$.
\end{proof}
\bibliographystyle{IEEEtran}
\bibliography{ref}
\end{document}